\documentclass[aps,prd,superscriptaddress,groupedaddress,12pt]{revtex4}

\usepackage{dcolumn}   % needed for some tables
\usepackage{bm}        % for math

\usepackage{amsmath}
\usepackage{amsfonts}
\usepackage{amssymb}
\usepackage{makeidx}
\usepackage{graphicx}
\usepackage{anysize}

\newcommand{\ba}{\begin{eqnarray}}
\newcommand{\ea}{\end{eqnarray}}
\newcommand{\be}{\begin{equation}}
\newcommand{\ee}{\end{equation}}

\usepackage[dvips]{color}

\begin{document}
\title{Using electron scattering to constrain the axial-vector form factor
\footnote{Proceedings of 16th International Workshop on Neutrino Factories and Future Neutrino Beam Facilities - NUFACT2014, 25 -30 August, 2014. 
University of Glasgow, United Kingdom }}

\author{R.~Gonz\'alez-Jim\'enez\footnote{Speaker}}
\affiliation{Departamento de F\'isica At\'omica, Molecular y Nuclear, Universidad de Sevilla, 41080 Sevilla, Spain}

\author{J.A.~Caballero}
\affiliation{Departamento de F\'isica At\'omica, Molecular y Nuclear, Universidad de Sevilla, 41080 Sevilla, Spain}

\author{T.W.~Donnelly}
\affiliation{Center for Theoretical Physics, Laboratory for Nuclear Science and Department of Physics,
Massachusetts Institute of Technology, Cambridge, Massachusetts 02139, USA}

\date{\today}

\begin{abstract}
We present an analysis of elastic and quasielastic parity-violating (PV) electron scattering processes. 
These reactions can help to constrain the weak neutral current form factors of the nucleon that play an 
essential role in the description of neutrino cross sections at intermediate energies.
We show that combining information from the analysis of elastic and quasielastic reactions the current 
knowledge on the strange and axial-vector form factors can be significantly improved.
\end{abstract}

\maketitle

\section{Introduction}

Motivated by the discovery of neutrino oscillations, recent years have been marked by a huge activity and new 
initiatives in both experimental and theoretical neutrino physics.
Most of the neutrino scattering experiments that have been proposed or recently carried out involve nuclear targets. 
Therefore, a good understanding of the neutrino-nucleus cross sections is essential in order to reduce the uncertainties 
in the determination of the oscillation parameters.
Many of these experiments (see \cite{Alvarez-Ruso14}), MiniBooNE, SciBooNE, Miner$\nu$a, NOMAD, K2K, T2K, 
have been designed to work
at the intermediate energy regime (from hundreds of MeV to a few GeV) where the quasielastic (QE) process is one of 
the dominant channels in the reaction mechanism. 
In this energy range, nucleon form factors play a fundamental role in the description of the cross section.
In this work we aim to show that parity-violating electron scattering can be used to study the form factors 
that enter in the weak neutral current of the nucleon.

The use of electrons as projectiles in comparison to neutrinos has important advantages:
i) electrons are easily produced, accelerated and detected, and ii) it is possible to produce monochromatic beams. 
In particular, the latter (monochromatic beams) allows one to have better control of the kinematics, 
since it is easier to estimate which specific channels are 
involved in the observed cross section (quasielastic, resonances, deep inelastic scattering, etc.).

\begin{figure}[htbp]
\centering
        \includegraphics[width=0.7\textwidth,angle=0]{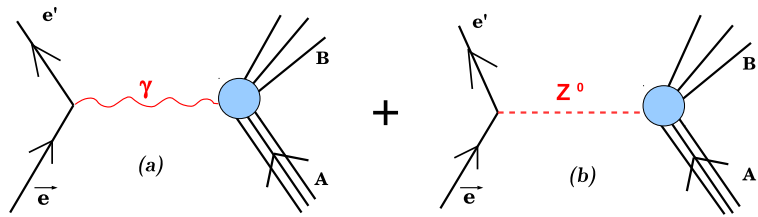}
        \caption{First-order Feynman diagrams for PV electron scattering: 
        (a) one photon exchanged, EM interaction, (b) one $Z^0$-boson exchanged, WNC interaction.}
        \label{fig:diagrams}
\end{figure}

In parity-violating electron scattering experiments a longitudinally polarized electron is scattered from a
nucleon (proton) or a nucleus, the electron being detected in the final state.
The Feynman diagrams describing the scattering process (in Born approximation) are shown in Fig.~\ref{fig:diagrams}. 
Although the electromagnetic (EM) interaction, mediated by the exchange of a virtual photon (diagram (a)), 
is dominant, the electron also interacts with the target through the weak neutral current (WNC) interaction, 
mediated by the exchange of a virtual $Z^0$ boson (diagram (b)).
Therefore, the cross section ($\sigma$) consists of the sum of three terms; the pure EM contribution, an
interference term between the EM and WNC currents and a purely WNC contribution:
\ba
 \sigma \propto |{\cal M}_{\gamma} + {\cal M}_Z|^2 = |{\cal M}_{\gamma}|^2 
 + 2{\cal R}e({\cal M}_{\gamma}^*{\cal M}_Z) + |{\cal M}_{Z}|^2\,,
\ea
where ${\cal M}_\gamma=j_\gamma^\mu J^\gamma_\mu$ with $j_\gamma^\mu$ ($J_\mu^\gamma$) the EM leptonic (hadronic) current. 
Similarly, ${\cal M}_Z=j_Z^\mu J^Z_\mu$, with $j_Z^\mu$ ($J_\mu^Z$) the WNC leptonic (hadronic) current.

The parity-violating asymmetry (${\cal A}^{PV}$) is defined as 
\ba
 {\cal A}^{PV} = \frac{\sigma^+ - \sigma^-}
			{\sigma^+ + \sigma^-}
		    = \frac{\sigma^{PV}}{\sigma^{PC}}\,,
		    \label{APV}
\ea
where $\sigma^{+/-}$ represents the cross section with positive/negative helicity of the incident electron. 
On the one hand, the denominator in the asymmetry is dominated by the EM contribution,
$\sigma^{PC} \propto |{\cal M}_{\gamma}|^2$,
that is, a parity conserving (PC) cross section. 
On the other hand, the first-order contribution in the numerator is the EM-WNC interference term, with the purely 
WNC contribution being several orders of magnitude smaller. Thus, 
$\sigma^{PV} \propto 2{\cal R}e\left({\cal M}_{\gamma}^*{\cal M}_Z\right)$, 
i.e., a parity-violating (PV) cross section.
Notice that the PV asymmetry is different from zero due exclusively to the presence of the weak interaction.
For this reason, the PV asymmetry can be used to study the different ingredients that enter in the weak neutral current. 
In particular, in this work we focus on the analysis of the WNC form factors of the nucleon, paying special attention 
to the axial-vector one.

We analyze two different processes, PV elastic electron-nucleon scattering (PVE), section~\ref{PVE}, and PV 
quasielastic electron-nucleus scattering (PVQE), section~\ref{PVQE}.
In the former, 
we have performed a statistical analysis of the full set of PVE asymmetry data 
(elastic electron scattering off proton~\cite{SAMPLE05,HAPPEX99,HAPPEXa,HAPPEXb,HAPPEXIII,PVA404,PVA405,PVA409,G005,G010,Qweak13} 
and $^4$He~\cite{HAPPEXb,HAPPEXHe}) 
providing estimates on the WNC form factors, in particular, on the electric (E) and magnetic (M) strange form factors 
of the nucleon ($G_{E,M}^{(s)}$) and on the axial-vector one ($G_A^{ep}$).
In section~\ref{PVQE} we present a brief discussion on the PVQE asymmetry and show how this observable could 
provide information on the WNC nucleon form factors that complements what is obtained from the elastic reaction. 
In particular, it is shown that this observable could help to constrain the isovector contribution in the 
axial-vector form factor~\cite{inclusive}.

\section{Parity-violating elastic electron-proton scattering}
\label{PVE}

After some algebra (see~\cite{Gonzalez-Jimenez13a} for details), 
the parity-violating elastic electron-proton asymmetry (${\cal A}^{PV}_{ep}$) can be written in the form:
\ba
{\cal A}^{PV}_{ep} = -\frac{{\cal A}_0}{2}  \left[
   \xi_V^p + \frac{G^{pn}}{G^{pp}}\xi_V^n
    +\frac{\xi_V^{(0)}\varepsilon G_E^p}{G^{pp}}G_E^{(s)} 
    + \frac{\xi_V^{(0)}\tau G_M^p}{G^{pp}} G_M^{(s)}
    - \frac{(1-4\sin^2\theta_W)\delta'G_M^p}{G^{pp}}G_A^{ep}
		     \right],\label{APVep}
\ea
where ${\cal A}_0$ is a function of the four-momentum transferred, $Q^2$, that determines the scale of the process. 
$\delta' = \sqrt{(1-\varepsilon^2)\tau(1+\tau)}$, being 
$\tau$ and $\varepsilon$ kinematic factors
(see ref.~\cite{Gonzalez-Jimenez13a} for explicit expressions).
$\theta_W$ is the weak mixing angle and the quantities $\xi_V$ are the WNC effective vector couplings.
Finally, $G_{E,M}^{p,n}$ are the EM form factors of the nucleon and the functions 
$G^{pp}=\varepsilon (G_E^p)^2 + \tau (G_M^p)^2$ and $G^{pn}=\varepsilon G_E^pG_E^n + \tau G_M^pG_M^n$ have been introduced. 

The axial-vector form factor can be decomposed in terms of a dominant isovector contribution ($G_A^{(3)}$) 
and two (octet, $G_A^{(8)}$, and strangeness, $G_A^{(s)}$) isoscalar contributions:
% by this way (see~\cite{}):
\ba
G_A^{ep} = \xi_A^{T=)} G_A^{(3)} + \xi_A^{T=0} G_A^{(8)} + \xi_A^{(s)} G_A^{(s)}\,.
\label{GAep}
\ea
Here the terms $\xi_A$ represent the WNC effective axial-vector couplings.  

In the case of elastic electron-$^4$He scattering the PV asymmetry can be written as (see~\cite{Musolf94}):
\ba
{\cal A}^{PV}_{eHe} = -\frac{{\cal A}_0}{2}
  \left[\xi_V^p+\xi_V^n + \frac{2\xi_V^{(0)}}{G_E^p+G_E^n}G_E^{(s)}\right]\,.\label{APVhe}
\ea

As observed, ${\cal A}^{PV}_{ep}$ (eq.~(\ref{APVep})) depends on the strange and axial-vector form factors, 
while ${\cal A}^{PV}_{eHe}$ (eq.~(\ref{APVhe})) depends on the electric strangeness. 
Consequently, a statistical analysis of the available experimental data on these observables provides information on 
the WNC nucleon structure. As already mentioned in the introduction, this is of great relevance for the analysis of
neutral-current (NC) neutrino scattering reactions at intermediate energies.

Some considerations are needed regarding the axial-vector form factor and the WNC effective couplings.
Corrections to the cross section from higher-order contributions, namely, radiative corrections (RC), are usually 
included in the WNC effective couplings ($\xi$) by modifying their {\it tree-level} values (see~\cite{Musolf94}).
However, a theoretical evaluation of these RC is not yet free from ambiguities. In fact, the contribution of RC 
is one of the main sources of uncertainties in the analysis of the PVE asymmetry.
In particular, contrary to neutrino scattering reactions where only the weak couplings are involved, in PV 
electron scattering, RC may play a very significant role in the description of the nucleon axial current and, 
consequently, in the axial-vector form factor.

At tree-level the value of the axial-vector form factor 
is\footnote{The values $G_A^{(3)}(0)\equiv g_A=1.27$ and 
$G_A^{(s)}(0)\equiv g_A^{(s)}=-0.08$~\cite{HERMES04,COMPASS07} has been assumed.}
$G_A^{tree}(Q^2) = -1.19 G^A(Q^2)$, where $G^A(Q^2)$ is a function of the four-momentum transferred.
According to the study of RC presented in ref.~\cite{Liu07}, the previous result should be modified to 
$G_A^{ep}(Q^2) = (-1.04\pm0.44) G^A(Q^2)$, its uncertainty  
being directly linked to the RC uncertainties. 
Thus, summarizing, RC may modify the tree-level value of $G_A^{ep}(Q^2)$ by more than 12\%, introducing also an
additional uncertainty of the order of 50\%.

In the present work we revisit the results from the statistical analysis of PVE asymmetry data 
presented in~\cite{Gonzalez-Jimenez14a}.
The EM form factors of the nucleon are assumed to be well under control, being described by the GKex 
prescription~\cite{Lomon01,Lomon02,Crawford10} (see~\cite{Gonzalez-Jimenez13a} for a detailed 
discussion on this topic).
Moreover, in order to include in our analysis data corresponding to a wide range of $Q^2$ values 
($0.02<|Q^2|<1$ (GeV/c)$^2$) the following $Q^2$-functional dependence of the strange and axial-vector form factors were used:
\ba
G_E^{(s)}(Q^2) = \rho_s \tau G_D^V(Q^2)\,,\,\,\,\,\, 
G_M^{(s)}(Q^2) = \mu_s G_D^V(Q^2)\,,\,\,\,\,\, 
G_A^{ep} = G_A^{ep}(0) G_D^A(Q^2) \, ,
\ea
where $G_D^V(Q^2) = (1+|Q^2|/M_V^2)^{-2}$ and $G_D^A(Q^2) = (1+|Q^2|/M_A^2)^{-2}$, with $M_V=0.84$ GeV/c$^2$ and $M_A=1.03$ GeV/c$^2$.
% Finally, it has been assumed that $\xi_V^{(0)} = -(1 + R_V^{(0)})$ with $R_V^{(0)} = -0.0123$ (see~\cite{Liu07}).

The analysis consists in fitting simultaneously the electric ($\rho_s$) and magnetic ($\mu_s$) strangeness parameters, 
the WNC effective couplings of the proton ($\xi_V^p$) and neutron ($\xi_V^n$) and the value of the axial-vector form 
factor at zero four-momentum transferred ($G_A^{ep}\equiv G_A^{ep}(0)$).
The linear dependence of the PV asymmetries (see eqs.~(\ref{APVep}) and (\ref{APVhe})) on the free parameters: 
$\rho_s$, $\mu_s$, $G_A^{ep}$, $\xi_V^p$ and $\xi_V^n$, simplifies importantly the problem since it is possible 
to perform an analytical $\chi^2$ fit (see~\cite{PDG12}).
Finally, notice that using $\xi_V^{p,n}$ and $G_A^{ep}$ as free parameters avoids introducing systematical errors 
linked to RC. Moreover, it also allows us to estimate the potential contributions from RC in the axial 
and vector sectors of the current by comparing the results from the fit to their tree-level values. 

\begin{figure}[htbp]
\centering
        \includegraphics[width=.5\textwidth,angle=270]{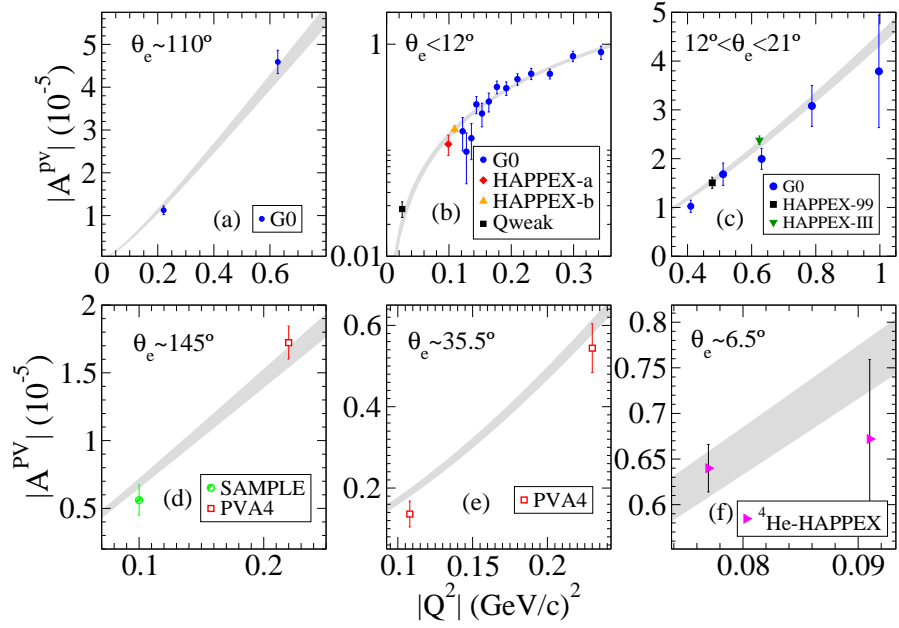}
        \caption{Full set of PV asymmetry data for elastic electron scattering compared with the prediction 
        from the $\chi^2$-fit (grey band). 
        The width of the band represents the theoretical uncertainty (1-$\sigma$ error). 
        The reduced-$\chi^2$ value is $1.30$.}
        \label{fig:comparison}
\end{figure}

In Fig.~\ref{fig:comparison} we present the comparison of the theoretical PV asymmetry and the experimental data. 
Each panel corresponds to a different scattering angle and the asymmetry is represented as a function of the 
four momentum transferred. 
In panel (f) the PV electron-$^4$He asymmetry is shown while the rest of panels correspond to the PV electron-proton 
asymmetry.  
The grey band represents the 1$\sigma$ error from the fit.
As observed, the agreement with data is rather good, particularly, at forward scattering angles (panels (b) and 
(c))\footnote{Notice that the high precision forward data, HAPPEX-III, HAPPEX-a, HAPPEX-b, dominate the fit.}.

\begin{figure}[htbp]
\centering
        \includegraphics[width=.3\textwidth,angle=270]{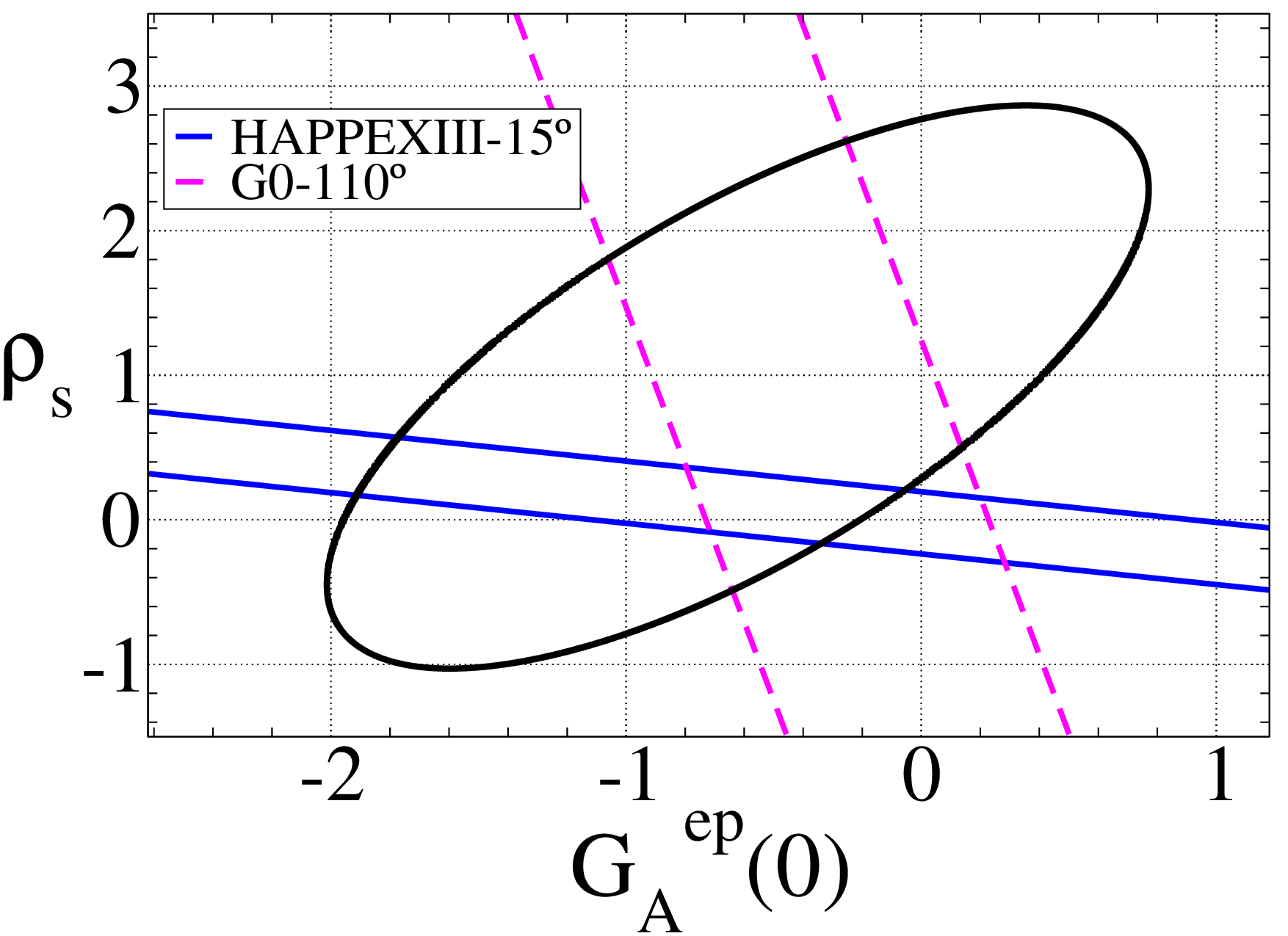}
        \includegraphics[width=.3\textwidth,angle=270]{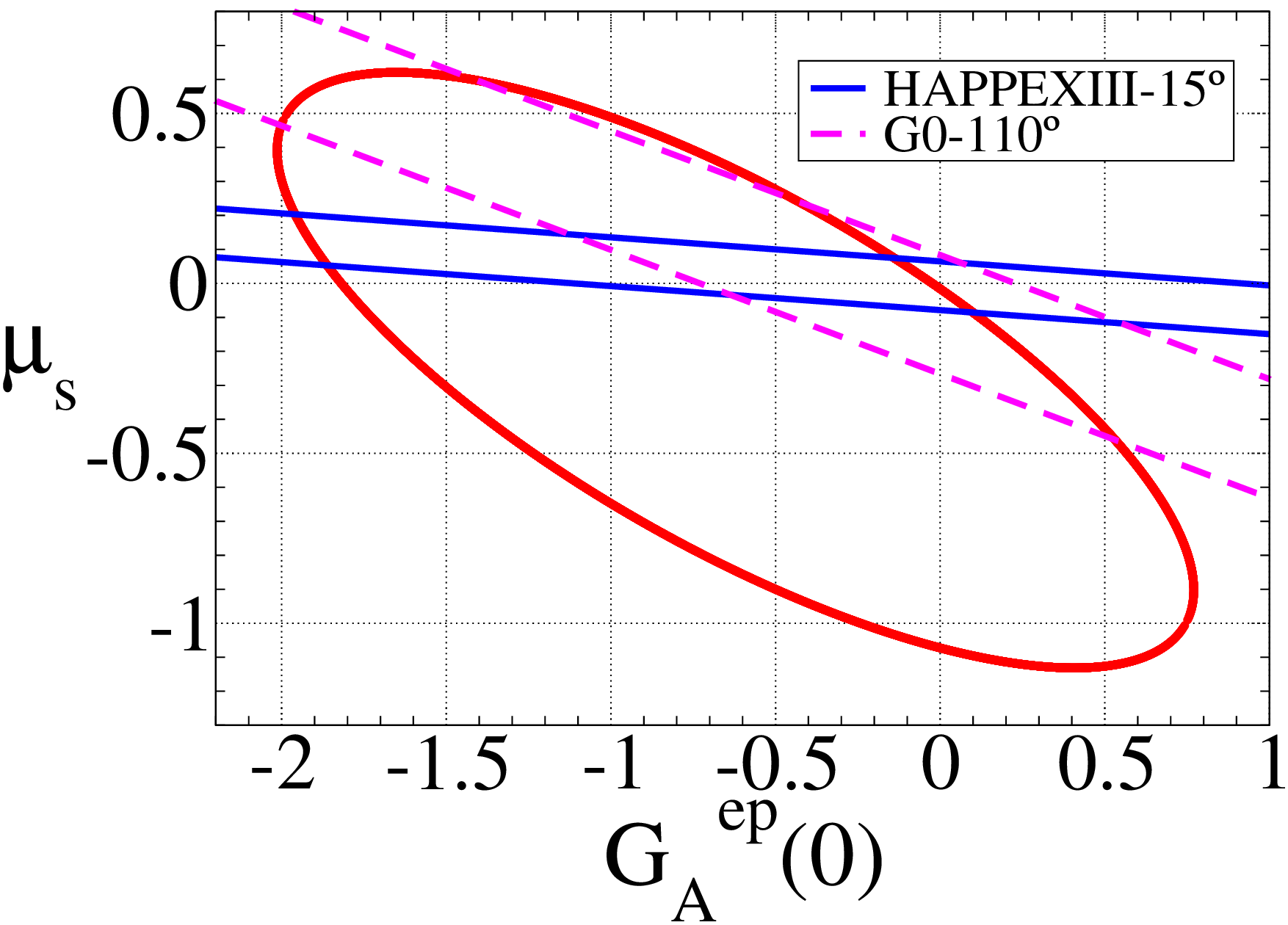}
        \caption{95\% confident level contours in the $[\rho_s-G_A^{ep}]$ (left) 
        and $[\mu_s-G_A^{ep}]$ (right) planes.
        The correlation coefficient for the couple $\rho_s\leftrightarrow G_A^{ep}$ is $0.711$ while 
        for $\mu_s\leftrightarrow G_A^{ep}$ is $-0.749$. 
        The straight lines represent the constraints from two experimental data at forward 
        (HAPPEXIII-15$^\circ$~\cite{HAPPEXIII}) and backward (G0-110$^\circ$~\cite{G010}) scattering angles.}
        \label{fig:ellipses}
\end{figure}

In Fig.~\ref{fig:ellipses} we represent the 95\% confident level ellipses for the parameters $\rho_s$, $\mu_s$ and $G_A^{ep}$.
The central values (point of maximum likelihood) and correlation coefficients for the full set of free parameters as 
well as other confident level ellipses have been presented in refs.~\cite{Gonzalez-Jimenez14a,Gonzalez-Jimenez14b}. 
Here we only discuss the results concerning the axial-vector form factor for which one gets $G_A^{ep}=-0.62\pm0.41$. 
Additionally, we have represented the constraints from two data, HAPPEXIII and G0-110$^\circ$, as examples of the 
two limit situations: forward and backward scattering.
In the $\rho_s-G_A^{ep}$ plane (left panel) the band corresponding to the backward data is almost vertical, that is, 
backward data provide essential information on the axial-vector form factor but not on $\rho_s$. The opposite occurs
at forward scattering, where data (horizontal band) basically constrain the electric strangeness parameter.
In the $\mu_s-G_A^{ep}$ plane (right panel) the situation is a somewhat different. 
Forward scattering data essentially constrain $\mu_s$ but not $G_A^{ep}$ while at backward kinematics results 
are sensitive to both $\mu_s$ and $G_A^{ep}$.

\section{Parity-violating quasielastic electron-nucleus scattering}
\label{PVQE}

In this section we study the inclusive parity-violating quasielastic electron-nucleus scattering process, $A(\vec{e},e')B$.  
Within the QE regime one considers that the longitudinally polarized electron interacts with only one nucleon in 
the target nucleus, with the struck nucleon being ejected from the nucleus. 
{\it Inclusive} refers to the fact that the only detected particle is the final electron.
In any other situation, for instance, in the exclusive case\footnote{In the exclusive reaction 
the ejected nucleon is detected in coincidence with the final electron, $A(\vec{e},e'N)B$.},
the pure EM responses
% , several orders of magnitud larger than the WNC ones, 
also contribute to the numerator of the PV asymmetry (\ref{APV}) making this observable useless to
study the weak neutral current (see~\cite{exclusive}).

\begin{figure}[htbp]
\centering
        \includegraphics[width=.35\textwidth,angle=270]{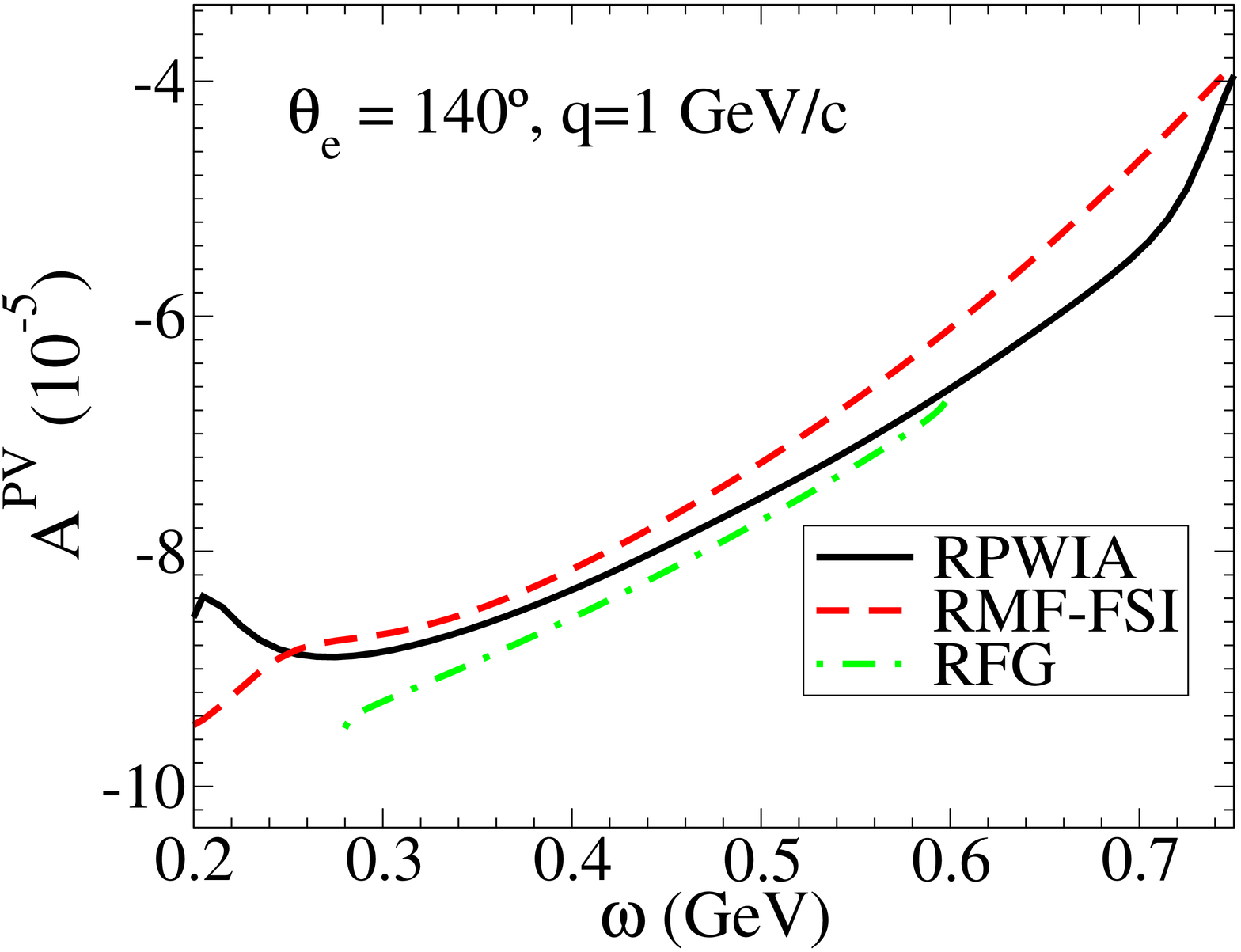}
	\includegraphics[width=.35\textwidth,angle=270]{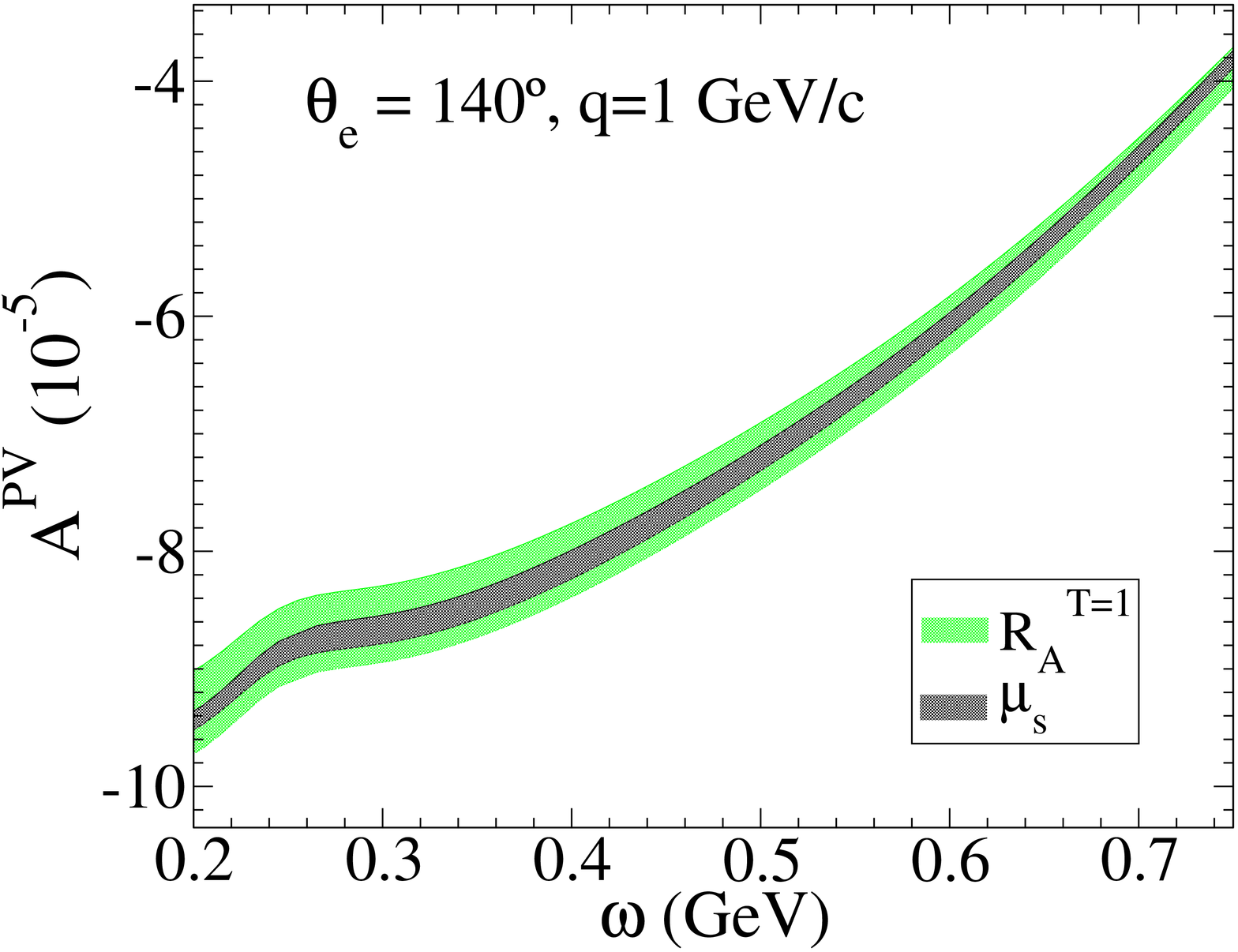}
        \caption{(Left panel) ${\cal A}_{QE}^{PV}$ computed with different models. 
        (Right panel) Effects of magnetic strangeness ($\mu_s$) and axial isovector radiative 
        corrections ($R_A^{T=1}$) on ${\cal A}_{QE}^{PV}$ computed with RMF-FSI model.
        In both panels the PVQE asymmetry is represented as a function of the energy transfer, $\omega$, 
        while the scattering angle and momentum transferred are fixed to $\theta_e=140^\circ$ and $q=1$ GeV/c.}
        \label{fig:APVQE}
\end{figure}

In the left panel of Fig.~\ref{fig:APVQE} we present the PVQE asymmetry computed with different nuclear models 
based on the impulse approximation:
\begin{itemize}
 \item {\it Relativistic Fermi gas (RFG)~\cite{Donnelly92}.} 
  The initial and final states of the struck nucleon are described as free-Dirac spinors. 
 \item {\it Relativistic plane-wave impulse approximation (RPWIA)~\cite{inclusive,exclusive}.}
  The bound nucleon wave function is a solution of the Relativistic Mean Field (RMF) Dirac equation 
  while the scattered nucleon is a Dirac plane wave.
 \item {\it Relativistic mean field with final-state interactions (RMF-FSI)~\cite{inclusive}.}
  The bound and scattered nucleon wave functions are solutions of the same RMF Dirac equation.
\end{itemize}
We see that the PVQE asymmetry is quite insensitive to the final state interactions (RPWIA {\it vs} RMF-FSI) 
and also to the description of the initial state of the nucleon (RPWIA {\it vs} RFG). 
In particular, in the region around the center of the QE-peak, $\omega\approx500$ MeV, 
the results of the three models deviate less than $\sim$ 7\%.

We have also studied the sensitivity of the PVQE asymmetry to nucleonic effects~\cite{inclusive}. 
In particular, at backward scattering angles the PVQE asymmetry shows special sensitivity to the description of the 
magnetic and axial-vector form factors. 
In the right panel of Fig.~\ref{fig:APVQE}, the effect of the magnetic strange parameter, $\mu_s$, is represented 
by the black band. 
The range of variation considered for $\mu_s$ is consistent with its prediction from the fit to the elastic data. 
The uncertainty in the isovector contribution of the axial-vector form factor leads to the green band shown in the figure. 

\section{Conclusions}
\label{conclusions}

We have presented a brief summary of some relevant results on PV elastic and quasielastic electron scattering 
published in previous studies~\cite{Gonzalez-Jimenez13a,Gonzalez-Jimenez14a,Gonzalez-Jimenez14b,inclusive,exclusive}.

In section~\ref{PVE} we have discussed the relevance of the PV elastic electron scattering asymmetry as an 
excellent tool to study the WNC form factors. 
We have shown the results of a statistical analysis ($\chi^2$ fit) of the full set of data using as free parameters 
the WNC effective vector couplings ($\xi_V^{p,n}$), the strangeness parameters ($\rho_s$ and $\mu_s$) and the value 
of the axial-vector form factor at zero $Q^2$ ($G_A^{ep}$).
An important result from the fit is the strong correlation existent between $\rho_s$, $\mu_s$ and $G_A^{ep}$.
Also, the fit provides an unexpectedly low value for $G_A^{ep}$ that could be understood as a signal of the importance
of RC effects in the axial-vector current (significantly higher than the current estimates). This result may also
indicate that alternative prescriptions for the $Q^2$ dependence of the strange and axial-vector form factors 
should be explored. Therefore, more studies on RC in the axial sector are essential before definite conclusions 
can be drawn on the vector strange form factors of the nucleon.

In section~\ref{PVQE} we have presented a brief discussion on the PV asymmetry for QE electron-nucleus scattering. 
Although additional uncertainties arise from the use of a complex nuclear target, we have   
shown that the PVQE asymmetry can provide nucleonic information that clearly complements the one attached to the 
PVE case. In particular, measurements of ${\cal A}^{PV}_{QE}$ at backward scattering angles could constrain the 
RC that enter in the isovector sector of the axial-vector form factor. This analysis, because of the strong 
correlation between the parameters, is essential in order to provide more accuracy estimates on the electric 
and magnetic strangeness contributions.


\begin{thebibliography}{99}
 \bibitem{Alvarez-Ruso14}
 L. Alvarez-Ruso, Y. Hayato and J. Nieves, New J. Phys. {\bf 16}, 075015 (2014) 
 %
 \bibitem{SAMPLE05}
 E. J. Beise, M. L. Pitt and D. T. Spayde. Prog. Part. Nucl. Phys., {\bf 54}, 289 (2005).
 %
 \bibitem{HAPPEX99}
 K. A. Aniol {\it et al.} [HAPPEX Collaboration], Phys. Rev. C., {\bf 69}, 065501 (2004).
 % 
 \bibitem{HAPPEXa}
 K. A. Aniol {\it et al.} [HAPPEX Collaboration], Phys. Lett. B., {\bf 635}, 275 (2006). 
 %
 \bibitem{HAPPEXb}
 K. A. Aniol {\it et al.} [HAPPEX Collaboration], Phys. Rev. Lett., {\bf 98}, 032301 (2007).
 %
 \bibitem{HAPPEXIII}
 Z. Ahmed {\it et al.} [HAPPEX Collaboration], Phys. Rev. Lett., {\bf 108}, 102001 (2012).
 %
 \bibitem{PVA404}
 F. E. Maas {\it et al.} [PVA4 Collaboration], Phys. Rev. Lett., {\bf 93}, 022002 (2004).
 % 
 \bibitem{PVA405}
 F. E. Maas {\it et al.} [PVA4 Collaboration], Phys. Rev. Lett., {\bf 94}, 152001 (2005).
 %
 \bibitem{PVA409}
 S. Baunack {\it et al.} [PVA4 Collaboration], Phys. Rev. Lett., {\bf 102}, 151803 (2009).
 %
 \bibitem{G005}
 D. S. Armstrong {\it et al.} [G0 Collaboration], Phys. Rev. Lett., {\bf 195}, 092001 (2005).
 %
 \bibitem{G010}
 D. Androi\'c {\it et al.} [G0 Collaboration], Phys. Rev. Lett., {\bf 195}, 092001 (2005).
 %
 \bibitem{Qweak13}
 D. Androic {\it et al.} [Qweak Collaboration], Phys. Rev. Lett., {\bf 111}, 141803 (2013).
 %
 \bibitem{HAPPEXHe}
 K. A. Aniol {\it et al.} [HAPPEX Collaboration], Phys. Rev. Lett., {\bf 96}, 022003 (2006).
  %
 \bibitem{inclusive}
 R. Gonz\'alez-Jim\'enez, J. A. Caballero and T. W. Donnelly,
 ``{\it Parity violation in quasielastic electron nucleus scattering within the relativistic impulse approximation}''. 
 In preparation (2015).
 %
 \bibitem{Gonzalez-Jimenez13a}
 R. Gonz\'alez-Jim\'enez, J. A. Caballero and T. W. Donnelly,
 Phys. Rep. {\bf 524}, 1 (2013).
 %
 \bibitem{Musolf94}
  M. J. Musolf and T. W. Donnelly, J. Dubach, S.J. Pollock, S. Kowalski and E.J. Beise, 
  Phys. Rep. {\bf 239}, 1 \& 2 (1994).
 %
 \bibitem{HERMES04}
 A. Airapetian {\it et al.}, [HERMES Collaboration],
 Phys. Rev. Lett. {\bf 92}, 012005 (2004).
 %
 \bibitem{COMPASS07}
 V. YU. Alexakhin {\it et al.}, [COMPASS Collaboration],
 Phys. Lett. B {\bf 647}, 8 (2007).
 %
 \bibitem{Liu07}
  J. Liu, R. D. McKeown and M. J. Ramsey-Musolf,
  Phys. Rev. C {\bf 76}, 025202 (2007).
 %
 \bibitem{Gonzalez-Jimenez14a}
 R. Gonz\'alez-Jim\'enez, J. A. Caballero and T. W. Donnelly,
 Phys. Rev. D. {\bf 90}, 033002 (2014).
 %
 \bibitem{Lomon01}
 E. L. Lomon, Phys. Rev. C {\bf 64}, 035204 (2001).
 %
 \bibitem{Lomon02}
 E. L. Lomon, Phys. Rev. C {\bf 66}, 045501 (2002).
 %
 \bibitem{Crawford10}
 C. Crawford {\it et al.}, Phys. Rev. C {\bf 82}, 045211 (2010).
 %
 \bibitem{PDG12}
 J. Beringer {\it et al.}, Phys. Rev. D {\bf 86}, 010001 (2012).
 %
 \bibitem{Gonzalez-Jimenez14b}
 O. Moreno, T. W. Donnelly, R. Gonz\'alez-Jim\'enez and J. A. Caballero, arXiv:1408.3511 [nucl-th] Aug (2014).
 %
 \bibitem{exclusive}
 R. Gonz\'alez-Jim\'enez, J. A. Caballero and T. W. Donnelly,
 ``{\it Parity violation and dynamical relativistic effects in $(\vec{e},e'N)$ reactions}''. In preparation (2015).
 %
 \bibitem{Donnelly92}
 T. W. Donnelly, M. J. Musolf, W. M. Alberico, M. B. Barbaro, A. De Pace and A. Molinari, Phys. Rev. A {\bf 541}, 525 (1992).
\end{thebibliography}
\end{document}